\documentclass[pra,superscriptaddress,twocolumn,floats,10pt,floatfix,nobibnotes]{revtex4-1}

\usepackage{amsmath}
\usepackage{amsfonts}
\usepackage{amssymb}
\usepackage{graphicx}
\usepackage{color}
\usepackage[colorlinks=true,citecolor=blue,linkcolor=blue,urlcolor=black]{hyperref}
\usepackage{times}
\usepackage{amstext}
\usepackage{latexsym}
\usepackage{comment}
\usepackage{bbm}

\begin{document}
\author{Feng Hu}
\email{f.hu.121214@gmail.com}
\affiliation{China Key laboratory of Specialty Fiber Optics and Optical Access Networks, Joint International Research Laboratory of Specialty Fiber Optics and Advanced Communication, Shanghai Institute for Advanced Communication and Data Science, Shanghai University, Shanghai 200444, China}
\affiliation{State Key Laboratory of Cryptology, P. O. Box 5159, Beijing, 100878, China}
\affiliation{Center for Quantum Computing, Peng Cheng Laboratory, Shenzhen 518000, China}
\author{Lucas Lamata}
\affiliation{Departamento de F\'isica At\'omica, Molecular y Nuclear, Universidad de Sevilla, 41080, Sevilla, Spain}
\affiliation{Department of Physical Chemistry, University of the Basque Country UPV/EHU, Apartado 644, 48080 Bilbao, Spain}
\author{Mikel Sanz}
\affiliation{Department of Physical Chemistry, University of the Basque Country UPV/EHU, Apartado 644, 48080 Bilbao, Spain}
\author{Xi Chen}
\affiliation{Department of Physical Chemistry, University of the Basque Country UPV/EHU, Apartado 644, 48080 Bilbao, Spain}
\affiliation{International Center of Quantum Artificial Intelligence for Science and Technology (QuArtist) and Department of Physics, Shanghai University, 200444 Shanghai, China}
\author{Xingyuan Chen}
\affiliation{State Key Laboratory of Cryptology, 100094 Beijing, China}
\author{Chao Wang}
\email{wangchao@staff.shu.edu.cn}
\affiliation{China Key laboratory of Specialty Fiber Optics and Optical Access Networks, Joint International Research Laboratory of Specialty Fiber Optics and Advanced Communication, Shanghai Institute for Advanced Communication and Data Science, Shanghai University, Shanghai 200444, China}
\affiliation{State Key Laboratory of Cryptology, P. O. Box 5159, Beijing, 100878, China}
\affiliation{Center for Quantum Computing, Peng Cheng Laboratory, Shenzhen 518000, China}
\author{ Enrique Solano}
\email{enr.solano@gmail.com}
\affiliation{Department of Physical Chemistry, University of the Basque Country UPV/EHU, Apartado 644, 48080 Bilbao, Spain}
\affiliation{International Center of Quantum Artificial Intelligence for Science and Technology (QuArtist) and Department of Physics, Shanghai University, 200444 Shanghai, China}
\affiliation{IKERBASQUE, Basque Foundation for Science, Mar\'{i}a D\'{i}az de Haro 3, 48013 Bilbao, Spain}

\begin{comment}
\email{lucas.lamata@ehu.eus}
\email{mikel.sanz@ehu.eus}
\end{comment}

\title{Quantum computing cryptography:\\
    Finding cryptographic Boolean functions with quantum annealing by a 2000 qubit D-wave quantum computer}

\begin{abstract}
As the building block in symmetric cryptography, designing Boolean functions satisfying multiple properties is an important problem in sequence ciphers, block ciphers, and hash functions. However, the search of $n$-variable Boolean functions fulfilling global cryptographic constraints is computationally hard due to the super-exponential size $\mathcal{O}(2^{2^n})$ of the space. Here, we introduce a codification of the cryptographically relevant constraints in the ground state of an Ising Hamiltonian, allowing us to naturally encode it in a quantum annealer, which seems to provide a quantum speedup. Additionally, we benchmark small $n$ cases in a D-Wave machine, showing its capacity of devising cryptographic Boolean functions with certain relevant properties. We have complemented it with local search and chain repair to improve the D-Wave quantum annealer performance related to the low connectivity. This work shows how to codify super-exponential cryptographic problems into quantum annealers and paves the way for reaching quantum supremacy.
\end{abstract}

\maketitle

%Keywords: Quantum Information; Quantum computation; Quantum annealing; Cryptography; Boolean functions

\section{Introduction}

Information security is of increasing concern involving in politics, military affairs, diplomacy, as well as in our daily life, where the security of communication systems plays a central role. Cryptography is important for the information security aiming at hiding the key information based on secure channels to defend from malicious parties.

The symmetric cryptosystem, including stream ciphers and block ciphers, is a typical way of implementing the encryption and decryption with the same key so that the high communication efficiency and security lead to wide applications in military defense, finance, and society. The performance of core cryptographic components that offer high security as the filter model, the combiner model, and S-box relies on the availability of Boolean functions~\cite{carlet2010boolean}. In fact, different cryptographic attacks~\cite{meier1988fast, courtois2003fast} require different properties such as, e.g., nonlinearity, balancedness, and correlation immunity.

However, there is a tradeoff among different criteria and it remains a challenge to achieve the best tradeoff to date~\cite{carlet2002larger, tang2013highly, picek2016cryptographic}. Resiliency and high nonlinearity are two important criteria proposed versus (fast) correlation attacks and best affine approximation (BAA) attacks~\cite{ding1991stability}. The properties of low-order resilient Boolean functions with high nonlinearity are important in stream ciphers. Although there exist several ways to find low-resilient and highly nonlinear Boolean functions, they may be limited by the search procedure of classical computers and the given functions with certain desired properties~\cite{zhang2017improving, carlet2002upper, pang2018construction}.

Although the size of the 1-resilient Boolean function with high nonlinearity is exponentially smaller than $2^{2^n}$, it is still difficult for classical computers in the sub-exponential space. It is necessary to find a new computing paradigm to explore the global properties of Boolean functions characterized in their exponential space.

Quantum annealing~\cite{finnila1994quantum} is an interesting alternative, and if the annealing progresses is sufficiently slow, natural quantum properties as quantum fluctuations and quantum tunneling effects can provide a quantum speedup in theory, at least for specific cases. More precisely, from the perspective of statistical theory, quantum-inspired systems can show a higher probability potentially to find the global optimum of multidimensional functions and can be seen as a global searching algorithm as compared with classical ones.

Google, Microsoft, IBM and a host of labs and start-ups are on the verge of a quantum technology breakthrough~\cite{boixo2016characterizing, castelvecchi2017quantum, pednault2017breaking}. Among them, D-Wave Systems, Inc.~\cite{dwavesys} is devoted to commercial quantum computing with prototypes based on the quantum annealing paradigm~\cite{dwavenature2011}, and mainly aimed at three categories of software applications and algorithms: Monte Carlo simulations, optimization, and machine learning. These include, among others, pattern recognition and anomaly detection, cyber security, image analysis, financial analysis, verification and validation~\cite{malley2018approach, mott2017solving, raouf2017prime, perdomo2012finding, courtlandd}. Additionally, we should also pay attention to the potential applications on encryption and decryption~\cite{chao2012impact, chao2016shaping, jiang2018quan}.

Nevertheless, up to now applications of the cryptography design by means of quantum computing have not been found. It is of great interest to consider the quantum annealing paradigm for Boolean function design as a way forward to implement the key cryptographic components design. The core quantum model underlying state-of-the-art D-Wave quantum annealing devices is the transverse-field Ising model, a basic model for mapping optimization problems onto a physical quantum annealer.

In this article, we propose, analyze, and experimentally implement a quantum annealing algorithm to design even-variable Boolean functions for use in cryptography. To this aim, we utilize the quantum theory to map the problem of designing Boolean functions with several criteria into the ground states of an Ising Hamiltonian. We consider $n$ to be even throughout the paper. Then, we obtain the Ising model ground states by quantum annealing experiments in the D-Wave cloud quantum computer, to illustrate a first demonstration of Boolean function design in a quantum computing way.

Up to now, no report has been produced on the cryptography designing and relevant issues by a quantum computer. First of all, the two typical applications of quantum computer reported earlier are the code-breaking technique and database searching method, both of which have nothing to do with the cryptography design. D-Wave Systems, Inc. realized a quantum annealing (QA) algorithm by using quantum tunneling effects, and this can be used to solve combinatorial optimization as well as some problems in the field of artificial intelligence. Moreover, even 2000-qubit problems were declared to be solved by D-wave 2X quantum computer. Up to now, however, none of the reports has announced to be able to handle the issues of cryptography designing. Thus, this article tries to apply the quantum annealing theory, based on quantum tunneling effects, to the cryptographic design and constructs the Boolean functions with multiple security criteria. This paper indicates the feasibility of applying D-Wave quantum computer to cryptography design. Finally, we have proposed that Quantum Computing Cryptography: A new age of critical applications of quantum computers, is coming.

This article considers the construction of Boolean functions with two important criteria, namely, nonlinearity and resiliency. We will consider the design problem as a search problem in an exponentially large solution space. All of our design procedures are based on the Walsh spectra to characterize the two criteria. Our research focuses on bent functions design and $m$-resilient functions design ($m \geq 1$) with high nonlinearity in small $n$ cases.

\begin{figure}[t]
\centering
\includegraphics[width = 3cm ]{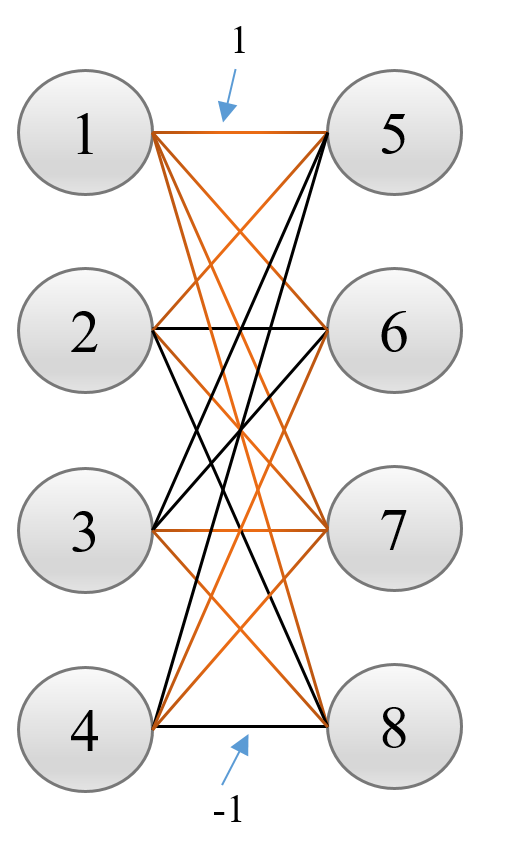}
\caption{Hardware architecture for designing 2-variable bent functions. Each circle $i$ denotes the qubit $\sigma_i$ and the lines between them give the strengths of the couplers between $\sigma_i$ and $\sigma_j$. The red line implies the coupler strength to be 1 while the black to be -1.}
\label{fig2bentfunIsing}
\end{figure}

\section{Boolean design scheme with a D-Wave quantum annealer\label{Boolean design}}

\subsection{Boolean functions}
\label{sec:boolfun}
An $n$-variable Boolean function $f(x)\in \mathfrak{B}_n$ is defined as the function from $\mathbb{F}^n_2$ to $\mathbb{F}_2$ and is generally represented by its algebraic normal form,

\begin{equation}
f(x)=f(x_1,...,x_n)=\bigoplus\limits_{u\in\mathbb{\mathbb{F}}^n_2}\lambda _u (\prod _{i=1}^{n}x_i^{u_i}),
\end{equation}

where $\lambda _u \in \mathbb{F}_2$, $u=(u_1,...,u_n)\in \mathbb{F}^n_2$, and the addition is denoted over $\mathbb{F}_2$.

 Any $f(x)$ could also be given in another form of a truth table as $[f(0,..,0), f(0,...,1),...,f(1,...,1)]$. The truth table consists of $2^n$ outputs of $f(x) \in \{0,1\}$, which is actually an exponential space with the size of $2^{2^n}$.

The algebraic degree ${\rm deg}(f)<2$ of $f(x)$ is denoted by ${\rm max}\{wt(u),\lambda_u \neq 0\}$, where $wt(u)$ is the Hamming weight of $u$. If  $wt(f)=2^{n-1}$, the function is balanced. The Boolean function with ${\rm deg}(f)<2$ , where ${\rm deg}(f)$ denotes the degree of $f$, is named ``affine function''. If the constant term equals to zero, it is called linear function, as below,

\begin{equation}
a\cdot x = a_1x_1 \oplus a_2x_2 \oplus ... \oplus a_nx_n
\end{equation}

where $a=(a_1,a_2,...,a_n)$, $x=(x_1,x_2...,x_n)\in \mathbb{F}^n_2$, and ``$\cdot$'' is the inner product of vectors $a$ and $x$.

The Walsh spectrum is an important concept to characterize different criteria. For any $a\cdot x \in \mathbb{F}_2$, the Walsh transform is the real-valued function over $\mathbb{F}^n_2$ defined as,

\begin{equation}
W_f(a)=\sum\limits_{x\in \mathbb{F}^n_2}(-1)^{f(x)+a\cdot x}
\end{equation}

In order to resist the best affine approximation~\cite{ding1991stabilityS} and fast correlation attacks~\cite{meier1988fastS}, the Hamming distance, which is characterized as the nonlinearity, to affine functions should be large enough. The nonlinearity of $f(x)$ is given by,

\begin{equation}
\label{nonlinearity}
nl(f)=2^{n-1}-\frac{1}{2}\cdot \max\limits_{a\in \mathbb{F}^n_2}|W_f(a)|
\end{equation}

Based on the Parseval's theorem~\cite{macwilliams1977theoryS}, if $n$ is even, $nl(f) \leq 2^{n-1}-2^{\frac {n}{2}-1}$, and even-variable Boolean functions with the nonlinearity achieving the upper bound are called bent functions.

Another criterion called correlation immunity is employed to characterize the ability of Boolean functions to resist the correlation attacks. For any $1\leq wt(a)\leq m$, if $W_f(a)=0$, $f$ is an $m$-order correlation-immunity function. And if $f$ with $m$-th order of correlation immunity is balanced, then we call it $m$-resilient function.

Generally, there are three ways to build Boolean functions: algebraic constructions, random search, and heuristics (and their combinations)~\cite{millan1998heuristicS, carlet2002largerS, tu2011conjectureS, picek2016cryptographicS}.

1) Algebraic constructions are provable in mathematics and can construct a class of Boolean functions with multiple criteria. Actually they are theoretically deterministic constructions to characterize a class of Boolean functions under specific conditions. Thus, they may result in certain classes of functions with similar properties, which are only a (small in most cases) subclass of functions relative to the size of $2^{2^n}$. Furthermore, it is not easy to find a good construction that could achieve the tradeoff among many criteria.

2) Random search is a relatively fast method to obtain many Boolean functions. Due to the vast size of the search space, it is not a sufficient way to find functions with excellent properties.

3) The last method lies in the intermediate position between the algebraic construction and random search. It usually divides the two-stage optimization as primary construction and secondary construction, where the outputs of the former are sent to be the inputs of the latter. Algebraic construction and heuristic technologies can be either the first or the second optimization (even both). However, from the perspective of heuristics as the size scales up, it is likely to get trapped in a local optimum and fail in scanning the whole exponential space.

Briefly speaking, known methods are unavailable to evaluate the global properties of Boolean functions, while designing the Boolean functions satisfying multiple criteria in the exponential-solution space is challenging. From the point of view of classical computers, it is hard to complete the optimization in an exponential search space. Therefore, here we employ the Ising model, a widely used model in quantum annealing algorithms, to globally characterize the class of Boolean functions with certain properties and solve it by means of the cloud quantum computer provided by D-Wave.

\subsection{Design procedure}

\begin{figure}[!htp]
\centering
\includegraphics[width = 6cm ]{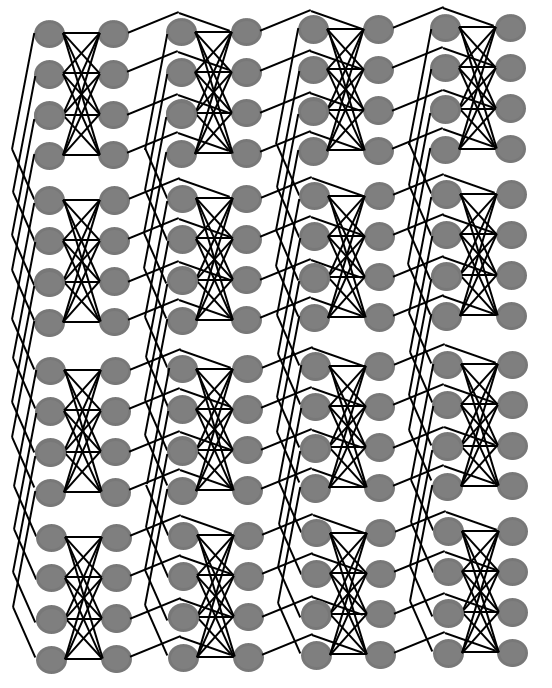}
\caption{Hardware architecture}
\label{Chimera_graph_1}
\end{figure}

An introduction to the design procedure is given below. However, there exist topological restrictions as shown in Fig.~\ref{Chimera_graph_1}, given that, in the D-Wave device, any qubits in two neighboring unit cells could only directly connect to the nearest neighbour qubits vertically or horizontally. This limits the scalability for large-scale systems such that we also introduce chains constructed with multiple physical qubits for representing one logical qubit for further scalability.

D-Wave provides a cloud quantum computer platform for connecting to via a local classical computer. In this work, the first step to design cryptographic Boolean functions in a quantum computing fashion consists of three steps, namely, 1) Recast the objective function into a set of polynomial terms and map it to an Ising model. 2) Design the hardware graph to solve the Ising model in a D-Wave quantum device. 3) Perform the quantum annealing computation via D-Wave cloud quantum computer and retrieve the corresponding Boolean functions with applications in cryptography.

Actually, we should first analyze the basic criterion, nonlinearity, in a logical way. For illustrative purposes, the case of 2-variable Boolean functions is given. Based on the Walsh spectra transformation introduced in Ref.~\cite{carlet2010boolean}, the Walsh spectra may be given in terms of a generalized matrix operation on the variants of the truth table.

Assuming that the truth table is $[\lambda_1,\lambda_2,\lambda_3,\lambda_4]$, where $\lambda_u \in \{0,1\}$, $u=1,2,3,4$, then a shift transformation $b_u=1-2\lambda_u \in \{-1,1\}$ maps the truth table onto $[b_1,b_2,b_3,b_4]$. Then, the Walsh spectrum can be given as,

\begin{equation}
\begin{split}
W_f(b)&=
\begin{bmatrix}
1 & 1 & 1 & 1\\ 1 & -1 & 1 & -1\\ 1 & 1 & -1 & -1\\ 1 & -1 & -1 & 1
\end{bmatrix}
\begin{bmatrix}
b_1\\ b_2\\ b_3\\ b_4
\end{bmatrix}
= \begin{bmatrix}
W_{f_1}(b)\\ W_{f_2}(b)\\ W_{f_3}(b)\\ W_{f_4}(b)
\end{bmatrix}
\end{split}
\end{equation}

Accordingly, the function is bent with the maximal nonlinearity if and only if $|W_{f_1}(b)|=|W_{f_2}(b)|=... =|W_{f_{2^n}}(b)|$. This is, the nonlinearity can become maximal if and only if all the absolute values of the Walsh spectrum are the same. In other words, the more uniform the absolute value of the Walsh spectrum, the higher the nonlinearity of the functions. Then, one can construct the Ising model for 2-variable bent functions in one unit cell as shown in Figure \ref{fig2bentfunIsing} based on the Hamiltonian as below,

\begin{equation}
\label{2bentfunIsing}
\begin{split}
&H_\textup{non}= \sigma _5(\sigma _1+\sigma _2+\sigma _3+\sigma _4)+\sigma _6(\sigma _1- \sigma _2+\sigma _3-\sigma _4)\\
&+ \sigma _7(\sigma _1+\sigma _2-\sigma _3-\sigma _4)+ \sigma _8(\sigma _1-\sigma _2-\sigma _3+\sigma _4).\\
\end{split}
\end{equation}

Based on the scalable structure given in Step 2, we employed the API provided by D-Wave to deliver the coefficients defined in the Ising model to the cloud quantum platform and can retrieve 1000 readouts at most in seconds.

%\begin{figure}[!htp]
%\centering
%\includegraphics[width = 4cm ]{unit_cell.png}
%\caption{The structure of scalable connections in the chimera graph of D-Wave machine. Any qubit in one 8-qubit unit cell can not connect to the qubits in the same column while the qubits in different unit cells can only connect correspondingly vertically or horizontally.}
%\label{Chimera_graph_2}
%\end{figure}

\section{Preliminary experiments and analysis}

Bent function (single criterion) design is a good benchmark choice as a step towards the goal of satisfying several appropriate criteria. In principle, the number of $n$-variable bent functions is exponential as shown in Table \ref{bentstatis}~\cite{hrbacek2014bent}.

\begin{table}[!htp]
\centering
\caption{$n$-variable Boolean functions}
\label{my-label}
\begin{tabular}{c|cccc}
\textit{n}         & 2 & 4 & 6 & 8 \\ \hline
Boolean functions  & $2^4$   & $2^{16}$  & $2^{64}$  & $2^{256}$  \\
bent functions     & $2^3$  & $\approx 2^{9.8}$  & $\approx 2^{32.3}$  & $\approx 2^{106.3}$  \\
relative frequency & $2^{-1}$  & $\approx 2^{-6.2}$  &$\approx 2^{-31.7}$  & $\approx 2^{-149.7}$
\end{tabular}
\label{bentstatis}
\end{table}

Based on the model constructed in Section \ref{Boolean design}, it can be generalized to $n$-variable bent functions that need $2^{n+1}$ logical qubits represented by $2^{2n-1}$ physical qubits, where the topological limitation  requires $2^{n-2}$ physical qubits to define a chain.

\begin{figure}[!htp]
\centering
\includegraphics[width = 3cm ]{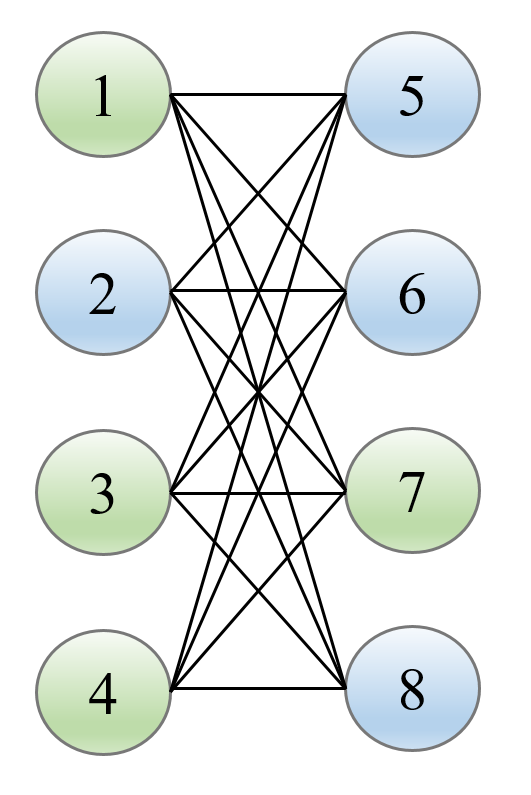}
\caption{One of the outcomes by the annealer on 2-variable bent case. If the green part represents state '1' and the blue part represents state '-1', it gives the corresponding truth table of the 2-variable Bent function in both sides. With more readouts from the machine, all the possible 2-variable bent functions can be obtained.}
\label{2-variable-graph}
\end{figure}

As an illustrative example with one unit cell consisting of eight working qubits, we obtain exactly eight bent functions out of 1, 000 readouts. We depict one of the cases in Fig.~\ref{2-variable-graph}, for which the left part ($[1,-1,1,1]$) and right part ($[-1,-1,1,-1]$) can be turned into two 2-variable bent functions as $[0,1,0,0]$ and $[1,1,0,1]$.

The cases of 4-variable and 6-variable bent functions are given in Section \ref{AdditionalExperiments}. What should be pointed out is that there exist two types of couplers in these cases: 1) the couplers between different chains (denoted as coupler strength), and 2) the couplers within the
chains (denoted as chain strength). With the comparison between the 4-variable case and 6-variable case, we find there exists a tradeoff between the coupler strength and the chain strength. Namely, the chains should be highly stable while the interactions between different chains should produce the optimal solutions.

As a consequence, we can find all the 4-variable bent functions, but can only find a reduced number of 6-variable bent functions due to several broken chains according to the considered regimes. Therefore, the quantum annealing protocol may end up with some suboptimal result.

Increasing the annealing time would not always work for different scenarios and some optimizations should be introduced based on the properties of D-Wave machine(the annealing time is fixed to 20 $\mu s$). We find that an additional local search algorithm may be useful for enhancement of our protocol. Here, we employ a basic greedy search algorithm, hill-climbing, to improve the results of the quantum annealer.

\emph{- Experiment 1: 6-variable bent functions}. In order to greatly utilize sub-optimal solutions, to select the ones with the nonlinearity more close to the maximum is important to achieve the tradeoff between the speed of postprocessing and the efficiency of the generation on bent functions. On the other hand, the readouts are given in order of increasing energies and in principle, the bent function is likely to occur with lower energy, such that applying the local search to the part of the results with lower energy can accelerate the design procedure.

Taking these two issues into account, Table \ref{6-var-opt} gives the number of distinct 6-variable bent functions out of 10 experiments. The coupler strength is fixed to 0.4 in order to maximize the performance to achieve the tradeoff between chain strength and coupler strength. The ``readouts'' column denotes the number of outputs once optimized by local search. The ``optimized range'' column denotes the nonlinearity of the sub-optimal Boolean functions produced by the D-Wave machine before the second optimization by local search. The ``numbers'' column denotes the number of 6-variable bent functions obtained with the D-Wave quantum processor and local search out of ten experiments. It is intuitive that the local search can greatly improve the performance of the quantum device.

\begin{table}[!htp]
\centering
\caption{6-variable bent functions design}
\label{6-var-opt}
\begin{tabular}{c|ccc}
coupler strength & readouts & optimized range & numbers \\ \hline
0.4                       & 50           & 27              & 134     \\
0.4                       & 50           & 26-27           & 296     \\
0.4                       & 50           & 25-27           & 664     \\
0.4                       & 100          & 27-28           & 208     \\
0.4                       & 100          & 26-27           & 354     \\
0.4                       & 100          & 25-27           & 864     \\
0.4                       & 200          & 27-28           & 218     \\
0.4                       & 200          & 26-27           & 558     \\
0.4                       & 200          & 25-27           & 1202
\end{tabular}
\end{table}

Furthermore, more readouts with the nonlinearity to be considered in the optimized range give more samples for the search and, accordingly, the system could obtain more bent functions. In other words, this indicates that the system can find many local points near to the optimal ones, while the classical one may get trapped in the local optimum, which is not a real bent function and can not be optimized directly by a local search.

{\emph{- Experiment 2: 4-variable $m$-resilient Boolean functions with high nonlinearity}. Due to the connectivity limit of the machine, we only consider the construction on  4-variable $m$-resilient Boolean functions with high nonlinearity. In theory, there is a tradeoff between the nonlinearity and resiliency order of a function~\cite{carlet2010boolean}.

Actually, the mapping of resiliency is a complete graph problem that cannot be directly merged into the nonlinearity. Although the number of logical qubits remains the same, the connectivity problem is more complicated such that the strength should be considered more carefully to ensure the stability of the chains. Relatively high coupler strength may lead to more broken chains and how to find the tradeoff between two different criteria should also be considered. Unfortunately, due to the complicated relationship between nonlinearity and resiliency, local search is unavailable in the case with several criteria.

To analyze it further, it is similar to add a new penalty term deduced by the truth table of the functions to the Ising model, which is a way to combine the nonlinearity and balancedness to characterize the final objective. We consider this according to the following four aspects: 1) The aim is to guarantee the function to be 1-resilient while maximizing the nonlinearity. 2) The working qubits for resiliency should relatively dominate the whole annealing procedure while the nonlinearity condition may also be fulfilled under the previous constraint. 3) Logically, normalization is necessary to balance the contributions of nonlinearity and resiliency to the objective functions. 4) Physically, the coupler strengths to stabilize the chains for different criteria should be considered separately. Thus, here we need another skill called chain repair to complete a majority vote to obtain the desired function. 

As a result, all the 4-variable 1-resilient Boolean functions with nonlinearity 4 have been found. Additionally, we observe that the chain repair can improve a relatively frustrating condition in the sense that the criteria require the regulation as accurate as possible on the chain strength and coupler strength. Therefore, with the chain repair, the parameter range may enable that some errors are eliminated or ignored as an error correction technique. In other words, to guarantee certain objectives with high loss at the cost of others with low loss is necessary to design the Boolean function satisfying multiple criteria.

\section{Topological restriction of the quantum device}
\label{sec:topology}

Figure~\ref{Chimera_graph_1} shows the partial $4 \times 4 \times 4$  chimera graph of the D-Wave 2000Q system ($16 \times 16 \times 4$) consisting of $4 \times 4$ unit cells including a $4 \times 2$ array. It intuitively illustrates the topological restrictions of the architecture, for which each qubit could only directly connect to the nearest neighbour qubits.

1) Given a single unit cell, it is divided into two parts: left and right. Each part contains 4 individual physical qubits without any connections while any two qubits in different parts can be connected with each other through the  so-called couplers.

2) Different unit cells connect only in two ways: the left part within each block connects vertically with other blocks, while the right part does it horizontally.

Obviously, each physical qubit could connect up to other neighboring six physical qubits (four qubits within the unit cell and up to two qubits connecting to the neighbouring unit cell). Users could set the weight of single qubit and coupler strength between two connected qubits. However, in most cases it is not sufficient to solve a practical problem with such a limited connectivity, thus a chain consisting of multiple physical qubits is introduced to represent one logical qubit to expand the connectivity.

In this way, the limited hardware architecture can be generalized to the condition of $n$-variable Boolean function as $n$ increases. Briefly, $n$-variable Boolean functions yield a $2^{n}\times 2^{n}$ coefficient matrix characterizing the Walsh spectra. For instance, the bent function requires in total $2^{n-2}\times 2^{n-2}$ unit cells and $2^{2n-1}$ physical qubits, which has been analyzed in the article.

\section{Additional Experiments\label{AdditionalExperiments}}
\label{ExpPreliminar}

As an extension of the previous constructions on the nonlinearity above, here we show how to implement the correlation immunity criterion, which is a similar model to the balancedness. With respect to the 2-variable Boolean function, the key is to find the corresponding coefficients if $m>0$ where $m$ is the total number of $'1'$ in the sequence ranging from $\{0,0\}$ to $\{1,1\}$, which is a one-to-one correspondence to the raw number. Thus, the cost function of 2-variable Boolean functions with the order of correlation immunity 1 is given by Eq. \ref{2resilient},

\begin{equation}
\begin{split}
\label{2resilient}
f_{\textup{corr}}= (b _1-b _2+b _3-b _4)^2+ (b _1+b _2-b _3-b _4)^2
\end{split}
\end{equation}

If we add the term $(b _1+b _2+b _3+b _4)^2$ defining the balancedness, the resiliency criterion can be mapped into the Ising model as,

\begin{equation}
\label{2resilientIsing}
H_{\textup{resi}}= \sigma _1\sigma _2 + \sigma _1\sigma _3 - \sigma _1\sigma _4- \sigma _2\sigma _3 + \sigma _2\sigma _4 + \sigma _3\sigma _4
\end{equation}

\textbf{\emph{Experiment 1: 2-variable bent functions}}
A 2-variable bent function design can be regarded as a proof of principle to check whether the protocol and device work as expected. The corresponding Hamiltonian reads,

\begin{equation}
\label{2bentfunIsing}
\begin{split}
H_\textup{non}= \left (\begin{bmatrix}
1 & 1 & 1 & 1\\
1 & -1 & 1 & -1\\
1 & 1 & -1 & -1\\
1 & -1 & -1 & 1
\end{bmatrix}
\begin{bmatrix}
\sigma_1\\
\sigma_2\\
\sigma_3\\
\sigma_4
\end{bmatrix} \right )^T \cdot
\begin{bmatrix}
\sigma_5\\
\sigma_6\\
\sigma_7\\
\sigma_8
\end{bmatrix}
\end{split}
\end{equation}

Based on the Ising model as the input of D-Wave, we attain 1, 000 readouts, among which eight exactly distinct bent solutions are found.

\textbf{\emph{Experiment 2: 4-variable bent functions}}

A more important issue is how to generalize the previous example to a high-variable case. Here, we denote the coefficient as $A$, and a new matrix, a $16 \times 16$ constant matrix, is easy to construct as,

\begin{equation}
H_{4-non}=\left (
\begin{bmatrix}
    A & A & A & A\\  A & -A & A & -A \\ A & A & -A & -A \\ A & -A & -A & A
\end{bmatrix}
\begin{bmatrix}
    \sigma_1\\ ... \\ ... \\ \sigma_{16}
\end{bmatrix}
\right )^T
\cdot
\begin{bmatrix}
    \sigma_{17}\\ ... \\ ... \\ \sigma_{32}
\end{bmatrix}
\end{equation}

It indicates that each logical qubit should interact with another 16 qubits. Thus, each chain indexing with the same color should be made up of 4 physical qubits to map the coefficients to the $4 \times 4$ arrays as in Fig \ref{4-variable-ppt}.

\begin{figure}[!htp]
\centering
\includegraphics[width = 6cm ]{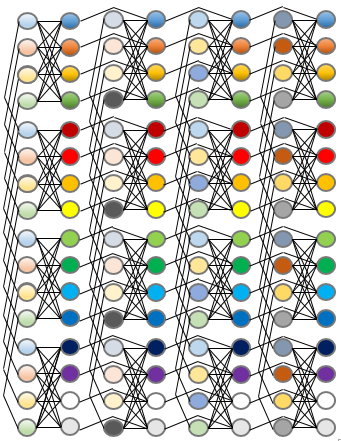}
\caption{The chain distributions for the 4-variable bent function design problem. }
\label{4-variable-ppt}
\end{figure}

The most important point is that there are two types of couplers in this case: 1) the couplers between different chains (denoted as coupler strength), and 2) the couplers within the chains (denoted as chain strength).

In order to clarify how they work, we fix the chain strength to $-1$ while varying the coupler strength from 2 to 0.1. The results are given in the Table \ref{4-var-bent-result}. We read 1, 000 solutions, each of which contains eight functions, to explore how many bent functions the system could generate and give the  average frequency out of 10 experiments.

\begin{table}[!htp]
\centering
\caption{The frequency of 4-variable bent functions}
\label{4-var-bent-result}
\begin{tabular}{c|cc}
coupler strength & bent frequency (\%) \\ \hline
2                          & 25.76                \\
1                         & 45.80                \\
0.5                       & 50.21                \\
0.25                      & 95.86                \\
0.1                       & 25.32
\end{tabular}
\end{table}

Here bent frequency denotes the frequency of bent functions obtained in the whole solution space derived from the D-Wave quantum annealer. That is, if the frequency is 95.86\%, it means the D-Wave machine can output about 96 bent functions out of 100 solutions. It also equals the success rate.

The exact number of 4-variable bent functions is 894, and there are many redundant ones. Thus, regardless of the randomness of the quantum platform, to improve the frequency can help us find more distinct bent functions. The table tells us that a suitable configuration could achieve a good performance. If the coupler strength is set to 0.25, all the results of the system are almost optimal with the same annealing time 20$\mu s$, which is enough for the machine to find the majority of optimal solutions. To analyze it further, we point out:

1) Large coupler strength will break the default chains such that the results are not the theoretical optimum all the time.

2) Similarly, small coupler strength cannot guarantee the interactions between different chains dominating the quantum annealing procedure such that the results may be affected by the chain strength to give suboptimal outputs.

3) There exists a tradeoff between the coupler strength and the chain strength in the sense that the chains should be highly stable while the interactions between different chains do still work to achieve the optimal solutions.

4) When it comes to large-scale systems, it can be predicted that small biases in the logical term will lead to nontrivial errors in the hardware graph and it may not be well corrected only by selecting the suitable chain strengths in long chain cases. The device may not always work consistently with the theoretical design due to the connectivity limitation of the hardware architecture.

\textbf{\emph{Experiment 3: 6-variable bent functions}}

As it comes to 6-variable case, it requires 2048 physical qubits representing 128 logical qubits, each of which is made up of 16 physical qubits as a chain, which is a more complicated condition compared with Exp. 2. By similar adjustments than with Exp. 2 on the device, we found it can hardly obtain 6-variable bent functions. In other words, too long and many chains will yield a significant amount of errors as compared with the 4-variable bent function design experiments. Although increasing the annealing time could find more optimal and suboptimal solutions, longer annealing times may be a trivial choice for optimization, and they also increase the risk of being affected by thermalization.

On the other hand, the advantage is that the running time of the quantum annealing processor does not grow as we extend the 2-variable case to the 6-variable case, although the qubit resources grow exponentially. Therefore, if some basic postprocessing technique could be introduced based on the properties of the quantum device, the performance will be better with the combination of the quantum annealing algorithm and classical algorithms.

\section{Optimization}
\label{sec:optianalysis}

Section \ref{ExpPreliminar} provided the experimental results of designing $n$-variable bent functions ($n=2,4,6$) by a quantum computer. Limited by the connectivity of the chimera graph, our protocol only works well on small cases (2 and 4-variable bent functions) and large case will cause many errors that could prevent obtaining suitable solutions. As a consequence, the chains are actually broken such that the qubits within the chain are not always aligned during the annealing. Here, we consider two improved strategies to optimize the results, based on the properties of the D-Wave quantum platform: 1) local search; 2) repairing the chain via majority vote~\cite{kristen2014errorS}.

As the evolutionary procedure of the quantum annealing is natural, each annealing procedure will give a number of optimal or suboptimal results including some  identical ones. Accordingly, the most simple way is to execute a number of experiments initialized with the same configuration to generate many different bent functions. But this approach is highly time-consuming, while optimization on the suboptimal outcomes offers a new feasible path to achieve the optimal solutions.

The local search algorithm is a simple but effective way to use the approximately optimal outcomes to obtain more desired functions. This article only employs a basic hill-climbing algorithm, to verify the advantage of introducing local search methods. Hill-climbing is a greedy search algorithm and always aims at the better solutions with multiple iterations in its local search field. Therefore, it can work well in the case of optimizing one criterion.

\textbf{\emph{Experiment 4: Employing local search to optimize 4-variable generation}}

Considering postprocessing, our objective becomes to obtain the suboptimal solutions via our original protocol and subsequently apply the hill-climbing algorithm to achieve the ones with the nonlinearity closer to the maximal nonlinearity. Additionally, this strategy can help to find more bent functions at once, to contribute to find all the 4-variable bent functions within several experiments.

\begin{table}[]
\centering
\caption{Average number of 4-variable bent functions before and after local search out of 10 experiments}
\label{4-var-bent-opt}
\begin{tabular}{c|cc}
Coupler strength & Initialized & Optimized \\ \hline
1                 & 85            & 164.8        \\
0.5               & 283.6         & 663.8        \\
0.25              & 812           & 814.6
\end{tabular}
\end{table}

As shown in Table~\ref{4-var-bent-opt}, with a simple test based on the relatively suitable coupler strength for the 4-variable bent functions, local search gives a significant improvement on the  relatively worse case and also slightly improves the better case. As a consequence, we could find all the 4-variable bent functions faster and it would also work in the more complicated case.

This means that the the device could provide suboptimal solutions near to the optimum within a few bit-flips, which may be more important in a large-scale system. Moreover, we also discuss in the article that the optimization procedure on the 6-variable bent functions can be significantly improved via a basic search algorithm.

\textbf{\emph{Experiment 5: Balanced 4-variable Boolean functions}}

We consider now balancedness as a typical case of resiliency. Due to the connectivity limit of the chimera architecture, here we only analyze the design scheme of balanced 4-variable Boolean functions as the basis for further optimization on resiliency.

The mapping of balancedness is actually a complete graph problem that cannot be directly transformed from the nonlinearity. Thus, we need to introduce another $4 \times 4$ array to design the balancedness criterion.

In this case, the interactions between different qubits are more complicated than the case of optimizing the nonlinearity, for which each qubit should interact with another fifteen qubits. Here, we construct a symmetric structure consisting of sixteen chains, each of them containing eight physical qubits.

If the coupler strength is set to 0.25, i.e., a more suitable value for optimizing the nonlinearity, it seems unfeasible to achieve balanced ones. This is because the chains constructed for the balancedness criterion are longer than for the nonlinearity, such that the chains are more unstable and the coupler strength should be accordingly smaller. As the coupler strength decreases further, we could find many balanced functions.

 The previous is an intermediate approach towards resiliency optimization due to the similar formalization of nonlinearity and resiliency. From the perspective of the Ising Hamiltonian, designing Boolean functions with several criteria seems that adding a penalty term to the initial Hamiltonian for improving one criterion requires more physical qubits although the number of logical ones remains the same. 

\textbf{\emph{Experiment 6: 4-variable $m$-resilient Boolean functions with high nonlinearity}}

We know that low-resilient functions must be balanced and Experiment 6 aims at combining all the penalty terms together to design the 4-variable $m$-resilient Boolean functions with the best known nonlinearity and in principle there are only 222 $m$-resilient functions in the solution space with the size of $2^{16}$. However, Exp. 5 shows that different criteria require different coupler strengths. Moreover, the chains optimal for the resiliency criterion may be more easily broken than the ones optimized for nonlinearity.

Therefore, the best tradeoff may be achieved at the cost of more stable chains. We analyze the effects between the two criteria as shown in Table~\ref{4-non-res} based on the connectivity of nonlinearity and resiliency.

\begin{table}[]
\centering
\caption{Average number of Boolean functions with $m$-th order of resiliency out of 10 experiments}
\label{4-non-res}
\begin{tabular}{cccc}
strength (N) & strength (R) & Initialized & Optimized \\ \hline
0.25                  & 0.125            & 104  & 113.2                              \\
0.5                  & 0.125          & 28.4 & 32                             \\
0.125                  & 0.125            & 150.8 & 157.6                             \\
0.05                  & 0.125          & 191.6 & 196
\end{tabular}
\end{table}

Here, the first and second columns give the coupler strengths for the nonlinearity (N) and resiliency (R), respectively. The ''Initialized'' column gives the number of $m$-resilient Boolean functions given by the D-Wave device, and the "Optimized" column gives the number of $m$-resilient functions after repairing the chain via majority vote.

It seems the improvement is not significant, but what should be pointed out is that the initialized solutions are based on all the solutions included in the architecture, where each one could give at least nine outcomes considering there exist broken chains while the majority vote only gives one solution. Therefore, not only we can get more optimums, but also the sufficiency of postprocessing has been improved.

Because the length of chains for resiliency is double than for nonlinearity, the former two cases in Table~\ref{4-non-res} imply that the resiliency does not dominate the annealing procedure. Thus, when we fix the coupler strength of resiliency to a suitable value, as the coupler strength of nonlinearity increases, the number of $m$-resilient functions decreases. In the latter two cases, the couplers of resiliency dominate the annealing procedure while the other parts do still work such that the machine can output more desired solutions.

Moreover, the nonlinearity of 4-variable $m$-resilient functions (for $m > 1$) is 0, because the high resiliency order will limit the nonlinearity. The coupler strength in the last case is smaller than the third case, such that one may find more $m$-resilient functions and the condition for the nonlinearity will not work well.  Thus, we point out that:

1) If the strength for the resiliency condition  is too large relative to the nonlinearity condition, the mapping of the nonlinearity may not work well.

2) The resiliency and nonlinearity conditions limit each other and optimizing the relative strength may help to obtain desired functions with specific properties, e.g., 4-variable 1-resilient functions with high nonlinearity.

3) Although the small coupler strength may produce more errors, it can also be a good choice to find the functions with similar properties (like high resiliency order) as a consequence of the sub-optimal results. For example, in this case we have found all the 2-resilient and 3-resilient Boolean functions simultaneously.

In small cases, the errors may lead to a small part of nonaligned qubits in a chain, for which the majority vote can simply obtain the corrected state of the qubits. Certainly, this will not always work, especially for more complicated cases. Thus, the majority vote could improve the performance near the critical point corresponding to the case relatively worse  to the best case. Moreover, how to select the suitable strength for majority vote is also important, as near the almost optimal boundary of the different criteria.

\section{Scalability}
\label{sec:scalability}

The Walsh spectra of Boolean functions with $n$ variables require $2^n$ logical qubits to characterize, which allows for a simplified model for nonlinearity, balancedness, and resiliency. In fact, to characterize the balancedness or resiliency one needs $2^n$ logical qubits and one requires extra $2^n$ qubits to represent the nonlinearity. Intuitively, the $m$-resilient Boolean function is balanced to reduce the search space with size of $2^{2^n}$ into the size of $\begin{pmatrix} 2^{n} \\ 2^{n-1} \end{pmatrix}$, in which the classical computer may find a subclass of optimal functions but wŒÆould fail in getting all the globally optimal ones.

As further mapped into the hardware architecture of  the D-Wave, the quantum annealing is a potential way to search in the globally exponential space to explore the global properties of the Boolean function. However, the connectivity limits the direct interactions between different logical qubits and the chain is constructed to implement the logical connectivity at the physical level. For example, for $n$-variable Boolean functions, the nonlinearity requires $2^n$ physical qubits as a chain to represent a logical qubit, such that in total $2^{2n-1}$ physical qubits are needed. Moreover, additional $2^{2n-1}$ physical qubits are arranged as a second part connected to the part for nonlinearity to design the Boolean functions satisfying more criteria such as balancedness and resiliency.

However, as $n$ grows up within the limited hardware architecture, more qubits would be turned into one chain as shown in Table \ref{qubitsofexp}. This will cause more errors in the annealing procedure and the majority of chains will be broken, resulting in suboptimal results. The adjustment of the strength of the qubits and couplers can correct some errors but will be in principle unavailable when it comes to a large-scale system.

\begin{table}[!htp]
\centering
\caption{Number of qubits required at the logical and physical level}
\label{qubitsofexp}
\begin{tabular}{c|ccccc}
\textit{}                        & 2  & 4    & 6       & 8        \\ \hline
Logical($2^{n+1}$)     & 8  & 32   & 128   & 512      \\
Physical($2^{2n-1}$)   & 8  & 128 & 2048 & $2^{15}$
\end{tabular}
\end{table}

Actually, the connectivity problem is related to the accuracy problem. The chain is unstable because the strength of the coupler in the chain is finite ($-1$). The experiments show that if the chain strength can be made stronger, the robustness could be better but correspondingly the logical interactions characterized by the physical chains may be weaker with respect to the chain.

Actually, the scheme gives a generalized model for $n$-variable Boolean functions (for even $n$) in theory, and, if a full-connectivity quantum computer with high fidelity is provided, the results can be better. With a full-connectivity and low-error-rate device,  it is possible to explore the global properties of Boolean functions in the exponential space, which is hard for classical computers. Therefore, the connectivity of the quantum annealer physical hardware is the key problem for the scalability.

\section{Discussion}

In summary, we have introduced a quantum annealing protocol to design even-variable Boolean functions with suitably designed criteria based on the D-Wave quantum computing platform. Quantum annealing can be seen as a new computing paradigm for cryptography design with the potential to explore the global properties of Boolean functions.

Based on the experiments on the 2048-qubit chimera hardware architecture, we implement 2, 4, and 6-variable bent function design, balanced 4-variable Boolean function design, and 4-variable $m$-resilient Boolean function design with high nonlinearity, respectively. One of the main problems is to achieve the tradeoff between the stability of the chains and the coupler strengths. Moreover, the suboptimal results provided by the machine allow the classical algorithm to improve outcomes and find the optimal ones. The local search was able to effectively optimize the approximately optimal solutions to generate different bent functions, where classical computers may easily be trapped in suboptimal results. The majority vote is a suitable way for the experiments involving multiple criteria to estimate that the best tradeoff between the coupler strength and the chain strength has been achieved, while it could improve the efficiency of function generating procedure as well. Additionally, as the system scales up, the total execution time almost remains the same, which is a main advantage of the D-Wave device compared to classical computers, although a large system may lead to more errors in our cases.

Furthermore, this is a novel quantum computing application and it is also a new way to design cryptographic keys. Additionally, the D-Wave device actually completes the search problem in an exponential space with dimension of up to $2^{2048}$ and outputs the optimal solutions, which shows the ability of global search compared to classical computing methods.

As a step forward towards a quantum advantage under these cases, one will need 512 logical qubits to design the 8-variable Boolean function with multiple criteria, for which classical computers may only find a small subclass of good solutions. Both new models to decrease the demands on qubit numbers and better-connectivity quantum devices are necessary in the future to breakthrough the bottleneck in cryptography by quantum computing.

\section*{Acknowledgements}
This work is supported by the grant of ``the Special Zone Project of National Defense Innovation'', the National Natural Science Foundation of China (61572304 and 61272096), and the Key Program of the National Natural Science Foundation of China (61332019), Open Research Fund of State Key Laboratory of Cryptology. We also acknowledge support from the program of Shanghai Municipal Science and Technology Commission (18010500400 and 18ZR1415500), the Shanghai Program for Eastern Scholar, Ram\'on y Cajal Grant RYC-2017-22482, Grant PGC2018-095113-B-I00 (MCIU/AEI/FEDER, UE), Basque Government IT986-16, the projects QMiCS (820505) and OpenSuperQ (820363) of the EU Flagship on Quantum Technologies, and the EU FET Open project Quromorphic. We acknowledge the use of D-Wave quantum computing facilities through Oak Ridge National Laboratory (ORNL).

\end{document}